%% file: main.tex
\documentclass[letterpaper,twocolumn,10pt]{article}
\usepackage{usenix2019_v3}

\usepackage{tikz}
\usepackage{amsmath}
\usepackage{booktabs}

\usepackage{xspace}
\usepackage{enumitem}

\usepackage[ruled,linesnumbered]{algorithm2e}
\usepackage{bbm}
\usepackage{colortbl}
\usepackage{xcolor}
\usepackage{pifont}
\usepackage{multirow}
\usepackage{url}
\usepackage{hyperref}

\usepackage[hang,flushmargin]{footmisc}

\input{macros}

\begin{document}

\date{}

\title{\sys: Harvesting Preemptible Cloud Resources for\\ Cost-Efficient Reinforcement Learning on LLMs}

\author{
\rm{Yongji Wu$^{\text{1}, *}$\enskip
    Xueshen Liu$^{\text{2},*,\text{\textdagger}}$ \enskip
    Haizhong Zheng$^{\text{4}}$ \enskip
    Juncheng Gu$^{\text{3}}$ \enskip} \\
\rm{Beidi Chen$^{\text{4}}$ \enskip
    Z. Morley Mao$^{\text{2}}$ \enskip
    Arvind Krishnamurthy$^{\text{3},\text{5}}$ \enskip
    Ion Stoica$^{\text{1}}$ \enskip}\\ \\
{$^{\text{1}}$UC Berkeley\enskip$^{\text{2}}$University of Michigan\enskip$^{\text{3}}$Google\enskip$^{\text{4}}$CMU\enskip$^{\text{5}}$University of Washington}
}

\maketitle

\input{abstract}

{\let\thefootnote\relax\footnote{
$^*$Yongji Wu and Xueshen Liu contributed equally. \\
\textsuperscript{\textdagger}Work was done when Xueshen Liu interned at Google.}}

\input{intro}
\input{background}
\input{overview}
\input{design}

\input{implementation}
\input{evaluation}
\input{discussion}

\input{related}
\input{conclusion}

\bibliographystyle{plain}
\bibliography{reference}

\input{appendix}

\end{document}

%% file: macros.tex
\newcommand{\sys}{RLBoost\xspace}

\newcommand{\rollouter}{rollout instance\xspace}
\newcommand{\rollouters}{rollout instances\xspace}
\newcommand{\trainers}{training cluster\xspace}

\newcommand{\disaggbl}{Disagg.BAL\xspace}

\newcommand{\cmark}{\textcolor{green!70!black}{\ding{51}}} 
\newcommand{\xmark}{\textcolor{red}{\ding{55}}}            

\def\Snospace~{\S{}}

\newcommand{\autorefsuffix}[2]{\hyperref[#1]{\autoref*{#1}#2}}

\DeclareMathOperator*{\argmax}{arg\,max}
\DeclareMathOperator*{\argmin}{arg\,min}

%% file: abstract.tex
\begin{abstract}
Reinforcement learning (RL) has become essential for unlocking advanced reasoning capabilities in large language models (LLMs). RL workflows involve interleaving rollout and training stages with fundamentally different resource requirements. Rollout typically dominates overall execution time, yet scales efficiently through multiple independent instances. In contrast, training requires tightly-coupled GPUs with full-mesh communication. Existing RL frameworks fall into two categories: co-located and disaggregated architectures. Co-located frameworks fail to address this resource tension by forcing both stages to share the same GPUs. Disaggregated architectures, without modifications of well-established RL algorithms, suffer from resource under-utilization. Meanwhile, preemptible GPU resources, i.e., spot instances on public clouds and spare capacity in production clusters, present significant cost-saving opportunities for accelerating RL workflows, if efficiently harvested for rollout.

In this paper, we present \sys, a framework for cost-efficient RL training that harvests preemptible GPU resources. 
Our key insight is that rollout's stateless and embarrassingly parallel nature aligns perfectly with preemptible and often fragmented resources. 
To efficiently utilize these resources despite frequent and unpredictable availability changes, \sys adopts a hybrid architecture with three key techniques: (1) adaptive rollout offload to dynamically adjust workloads on the reserved (on-demand) cluster, (2) pull-based weight transfer that quickly provisions newly available instances, and (3) token-level response collection and migration for efficient preemption handling and continuous load balancing. Extensive experiments show \sys increases training throughput by 1.51x-1.97x while improving cost efficiency by 28\%-49\% compared to using only on-demand GPU resources. \sys is open-sourced at \url{https://github.com/Terra-Flux/PolyRL}.

\end{abstract}

%% file: intro.tex
\section{Introduction}\label{sec:intro}

\begin{figure}
\centering
\includegraphics[width=0.99\linewidth]{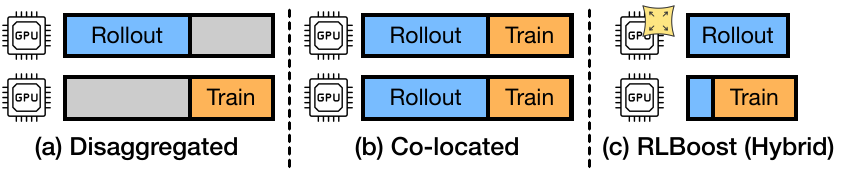}
\caption{Architectures for on-policy RL frameworks.}
\label{fig:arch_compare}
\end{figure}

Reinforcement learning (RL) post-training has become the key enabler in unlocking advanced reasoning capabilities for modern large language models (LLMs). RL not only empowers state-of-the-art LLMs like Claude 4~\cite{claude4} and Grok 4~\cite{grok4} to achieve leading performance in mathematics, coding, and tool use, but also enables smaller, more efficient models to reach or even surpass the performance of much larger LLMs on specialized tasks~\cite{zheng2025deepresearcher,prabhakar2025omniscience,wu2025agentic,bamba2025xrpo, zhou2025sweet}.

Unlike traditional pre-training, the RL workflow is mainly composed of two interdependent stages: rollout and training. In the rollout stage, input prompts are fed to inference engines, e.g., vLLM~\cite{kwon2023efficient}, to generate a batch of responses. The responses are then used in the training stage to derive reward signals, compute loss, and update model weights. The updated model is subsequently transferred to the inference engines for the rollout stage of the next iteration.

Existing RL frameworks can be divided into two categories: disaggregated and co-located. In disaggregated frameworks, rollout and training stages are assigned to separate sets of GPUs. They either struggle with resource under-utilization (\autoref{fig:arch_compare}(a)) due to bubbles caused by stage dependency~\cite{shen2024nemo,hu2024openrlhf}, or sacrifice model accuracy by using asynchronous (off-policy) algorithms~\cite{zhong2025streamrl,fu2025areal,zheng2025prosperity, han2025asyncflow} to relax the stage dependency.

To maximize resource utilization under the well-established synchronous (on-policy) RL algorithms, co-located RL frameworks are proposed~\cite{sheng2025hybridflow,zhong2024optimizing,mei2024realhf}, where training and rollout stages are on the same set of GPUs. The two stages time-share the GPUs, with each GPU alternating between rollout and training, avoiding idle GPU cycles. However, the two stages exhibit fundamentally different resource requirements. In terms of resource types, the rollout stage partitions available GPUs into multiple independent rollout instances, using tensor parallelism within each instance and requiring no communication across instances. In contrast, the training stage generally employs fully sharded data parallelism (FSDP)~\cite{zhao2023fsdp} and/or 3D parallelism across all available GPUs, involving extensive full-mesh communication between GPUs. In terms of resource quantities, the rollout stage scales efficiently by spawning more independent rollout instances and substantially benefits from allocating more GPUs than the training stage, as generation takes up to 90\% of overall RL time under the co-located setting~\cite{he2025history}. 

How can we reconcile this fundamental resource tension under synchronous algorithms without compromising system efficiency or incurring prohibitive monetary costs? Public clouds and private production GPU clusters typically offer their excess capacity in the form of preemptible resources. Spot instances on public clouds provide considerable cost savings (up to 90\% cheaper)~\cite{thorpe2023bamboo}, while production clusters generally have unused GPUs reserved for online workloads~\cite{duan2024parcae,newell2021ras}. These instances can be preempted at any time by the infrastructure provider. Moreover, these spare GPU resources often suffer from fragmentation at multiple levels, leading to significant communication overhead. At the node level, available GPUs may spread across many nodes, with each node already partially occupied~\cite{weng2023beware,gpufragmentation}. At the cluster level, available nodes may be topologically scattered across different racks or pods, causing traffic to cross spine and core switches~\cite{rajasekaran2024cassini,clusterfragmentation}. These preemptible and fragmented resources, while poorly suited for training, align well with the rollout stage's embarrassingly parallel and stateless nature.

Our insight is that through a \textit{hybrid} architecture, we can harvest preemptible resources for high throughput and cost-efficient RL on LLMs. Under the hybrid architecture shown in \autoref{fig:arch_compare}(c), the reserved training cluster performs both training and rollout but opportunistically outsources part of the rollout workload to available preemptible rollout instances.

Still, to efficiently harvest these preemptible resources, there are several key challenges. First, how can we adapt the workloads on the training cluster to dynamic preemptible resource availability? 
Second, when a new preemptible instance becomes available, how can we quickly provision it with the latest model weights for it to begin rollout, while minimizing progress loss when an instance is preempted?
Third, how can we balance the load across rollout instances? The output length of rollout requests in RL exhibits high non-determinism~\cite{yu2025dapo}, which is further complicated by instance elasticity.  Without careful scheduling, tail requests cause severe load concentration on a small subset of instances.

To address these challenges, we propose \sys, an RL framework with a hybrid architecture that harvests preemptible resources. To adapt to dynamic resource availability, \sys employs an adaptive rollout offload mechanism. At each step, the training cluster starts from a "seeding" stage, where it is temporarily repurposed for rollout. During this stage, it pre-computes a part of rollout responses that serve as "seeds" for remote rollout instances to continue, before switching to training mode and overlapping with remote rollout through dynamic micro-batch pipelining. \sys adaptively tunes the seeding time window based on current workloads and preemptible instance availability.

To quickly provision weights to newly available instances, we decouple weight transfer logic from the training and inference frameworks. We design a pull-based transfer agent that asynchronously transfers model weights, enabling new instances to join and contribute to rollout at any point during a training step.

To minimize preemption overheads and enable fine-grained load balancing, \sys collects rollout results at token granularity rather than request level, allowing flexible request migration between rollout instances at any point without progress loss. Building on this token-level stream redirection mechanism, \sys incorporates a real-time load balancer that continuously monitors queue depths across rollout instances and redistributes in-flight requests.

We evaluate \sys using H100 GPU instances from a public cloud. Extensive evaluations ranging from 8B to 32B models with various spot instance traces show that \sys increases overall RL training throughput by 1.51x-1.97x while improving cost efficiency, i.e., the total tokens trained with the same monetary budget, by 28-49\%.

In summary, we make the following contributions:
\begin{itemize}[leftmargin=*]
    \item We identify the fundamental resource tension between the rollout and training stages in RL workflow, and propose a hybrid architecture to harvest preemptible resources for high throughput and cost-efficient RL.
    \item We design an adaptive rollout offload mechanism to dynamically adapt the training cluster's workloads to real-time resource availability, while adhering to well-established synchronous (on-policy) RL algorithms.
    \item We develop pull-based weight transfer to quickly provision weights to new instances, complemented by token-level response collection and migration to handle preemptions.
    \item We conduct extensive experiments to evaluate \sys and demonstrate its performance and cost efficiency against state-of-the-art RL frameworks. 
\end{itemize}

%% file: background.tex
\section{Background}

\label{sec:background}

\begin{figure}
\centering
\includegraphics[width=0.95\linewidth]{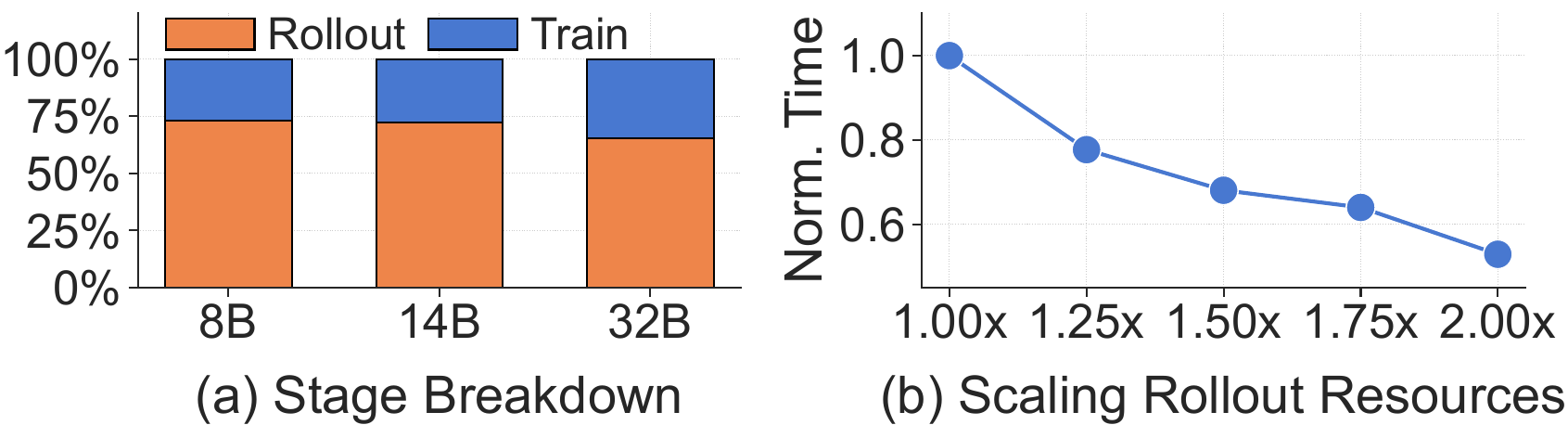}
\caption{The rollout stage dominates an RL step in the co-located architecture, yet it efficiently scales with more GPU resources as each rollout instance operates independently.}
\label{fig:background_gen_time}
\end{figure}

\subsection{Reinforcement Learning for LLMs}
Reinforcement Learning (RL) is a technique that predates LLMs but has emerged as the predominant paradigm for fine-tuning LLMs during post-training, aligning them with human preferences~\cite{ouyang2022training} and enhancing their performance on tasks requiring complex reasoning~\cite{guo2025deepseek,shao2024deepseekmath}. In a typical RL workflow, the process begins with a pre-trained base model that serves as the agent model to be optimized. The agent learns to take a set of actions given an input and receives different rewards based on the actions it takes. In the context of LLMs, the inputs are initial prompts that present tasks for the model to solve. The model takes actions by generating tokens autoregressively, where each generated token constitutes an individual action. The agent LLM's behavior is optimized by training it to learn action sequences that maximize the expected reward.

Although there are a variety of RL algorithms, e.g., PPO~\cite{schulman2017proximal} and GRPO~\cite{guo2025deepseek}, they all revolve around two main stages within each RL step: rollout and training. Each step begins with the rollout stage, where the LLM agent takes actions by processing a batch of prompts and generating a single or a group of responses per prompt, similar to traditional LLM inference. Upon completion, each generated response forms a training sample.
    
In the training stage, a reward~\cite{guo2025deepseek} is computed for each response (i.e., training sample) to derive the loss function for model updates.  The reward typically comes from a rule-based verifiable function, such as a binary signal indicating whether the response successfully completes a coding task or correctly answers a mathematical question. Alternatively, the reward can be derived from another LLM, referred to as a reward model. RL algorithms may also employ critic models or reference models to provide additional loss signals. However, these auxiliary models remain frozen during RL training and generally introduce insignificant computational overhead.  Mainstream LLMs are predominantly trained with synchronous (on-policy) RL algorithms~\cite{guo2025deepseek,shao2024deepseekmath,zheng2025act, yan2025oppo}. After the agent LLM is updated, the new model weights are immediately used in the next rollout stage, ensuring that responses are always generated using the latest version of the model.

\subsection{RL Frameworks}

\begin{table}[t]
\centering
\caption{Overview of existing RL frameworks for LLMs.}
\label{tab:systems_compare}
\resizebox{0.99\columnwidth}{!}{
\begin{tabular}{l|p{1.8cm} p{1.8cm} p{1.8cm}}
\toprule
\textbf{Systems} & \textbf{On-policy Optimized} & \textbf{Resource Decoupling} & \textbf{Preemptible Resources}\\
\midrule
veRL~\cite{sheng2025hybridflow}      & \cmark & \xmark & \xmark \\
StreamRL~\cite{zhong2025streamrl}  & \cmark & \cmark & \xmark \\
AReaL~\cite{fu2025areal}     & \xmark & \cmark & \xmark \\
AsyncFlow~\cite{han2025asyncflow} & \xmark & \cmark & \xmark \\
RhymeRL~\cite{he2025history} & \xmark & \cmark & \xmark \\
\rowcolor{gray!15} 
\sys      & \cmark & \cmark & \cmark \\
\bottomrule
\end{tabular}
}
\end{table}

\label{sec:background_rl_frameworks}
Existing RL frameworks can be categorized into two architectures: co-located and disaggregated. Early RL frameworks adopted the disaggregated architecture~\cite{shen2024nemo,hu2024openrlhf} to effectively reuse existing system infrastructures. The training stage is deployed on one set of GPUs using frameworks such as Megatron-LM~\cite{shoeybi2019megatron}, while the rollout stage is deployed with another set of GPUs using frameworks like vLLM~\cite{kwon2023efficient}. At each step, weights are first transmitted from the training workers to the rollout workers, after which the rollout and training stages execute sequentially. At any given time, one set of GPUs remains idle while waiting for the other set to complete its stage. There are some recent disaggregated frameworks that improve system efficiency by optimizing for asynchronous (off-policy) RL algorithms~\cite{zhong2025streamrl,fu2025areal,han2025asyncflow,he2025history}.

To address the resource utilization issue under the widely-adopted synchronous (on-policy) RL algorithms, the co-located architecture is developed~\cite{mei2024realhf,sheng2025hybridflow}, which switches between rollout and training on the same set of GPUs. However, there is a fundamental mismatch in resource requirements between the two stages as described in \autoref{sec:intro}. \autoref{fig:background_gen_time}(a) presents the step time breakdown for training Qwen3~\cite{qwen3technicalreport} models using the co-located veRL framework, with experimental details in \autoref{sec:eval_overall}. Rollout accounts for up to 73\% of the overall time, yet it can be easily accelerated with more GPU resources, as shown in \autoref{fig:background_gen_time}(b).

We compare existing RL frameworks in \autoref{tab:systems_compare}. None can leverage preemptible resources and adapt to dynamic resource availability, whether they are disaggregated or co-located.

%% file: overview.tex
\section{Overview}

\begin{figure}
\centering
\includegraphics[width=0.9\linewidth]{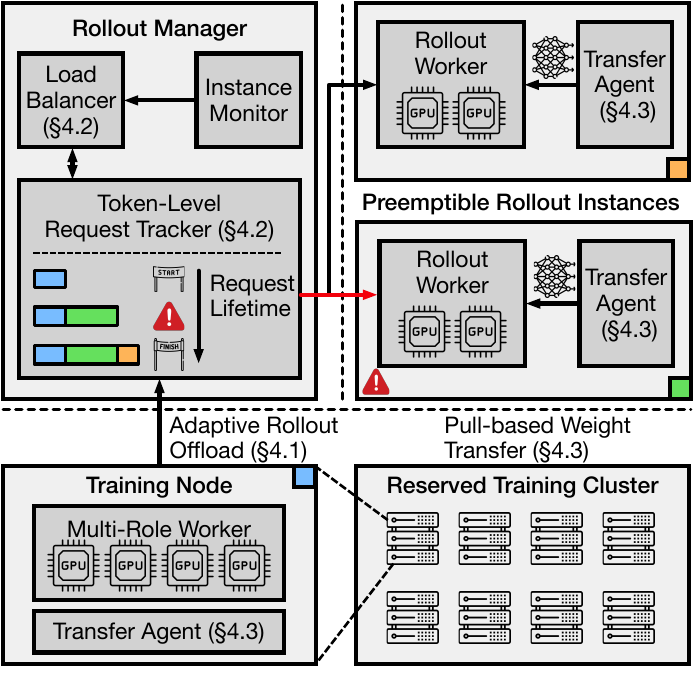}
\caption{System overview of \sys.}
\label{fig:overview}
\end{figure}

\sys is a hybrid RL framework that leverages preemptible instances for high-throughput and cost-efficient RL training. We present the major components of \sys in \autoref{fig:overview}. \sys employs a fixed (reserved) \trainers, as the training stage requires tightly coupled GPUs, and frequent preemptions would incur significant checkpoint-restart overhead. We note that existing fault-tolerant training techniques~\cite{wan2025robust,he2023unicron,wang2023gemini} are orthogonal to our work, and \sys can directly benefit from them. \sys also utilizes an elastic pool of preemptible GPU instances to offload rollout workloads from the training cluster, where instances can be dynamically allocated or preempted at any time. We refer to these preemptible instances dedicated to rollout offload as \textit{(remote) \rollouters}. These instances can be located either in the same datacenter or cloud region as the \trainers, or distributed across different datacenters or cloud regions. As shown in \autoref{fig:background_gen_time}, rollout typically consumes the majority of step time; therefore, offloading it to more affordable preemptible resources can significantly increase throughput while reducing monetary costs.

The core component connecting the \trainers and \rollouters is the rollout manager. It monitors the health of each \rollouter, handles preemptions, and launches rollout workers when new instances become available. 

In every step, the rollout manager sends a part of the rollout requests to \rollouters on behalf of the \trainers. It continuously tracks the status of each request and collects the responses in token granularity. A load balancer distributes requests across \rollouters and redirects in-flight requests upon load variations or instance preemptions.

To adapt the amount of workload offloaded from the training cluster to dynamic rollout instance availability and balance remote and local execution, \sys employs multi-role workers on the training nodes, which can be temporarily re-purposed for rollout at the beginning of each RL step.
While the \rollouters are receiving weights and generating the first stream of responses, the \trainers would handle rollout requests within a specific time window. Such mechanism enables the \trainers to "seed" a part of the responses for remote \rollouters to continue the work. To enable a new \rollouter to join and participate in the rollout at any time during a step, \sys decouples the weight transfer logic from the training and rollout workers into dedicated transfer agents. The agents send and receive weights asynchronously while the training node is occupied with either seeding rollout or training tasks. 

%% file: design.tex
\section{Design}
\label{sec:design}

\subsection{Adaptive Rollout Offload}
\label{sec:design_response_seeding}

\begin{figure}
\centering
\includegraphics[width=0.95\linewidth]{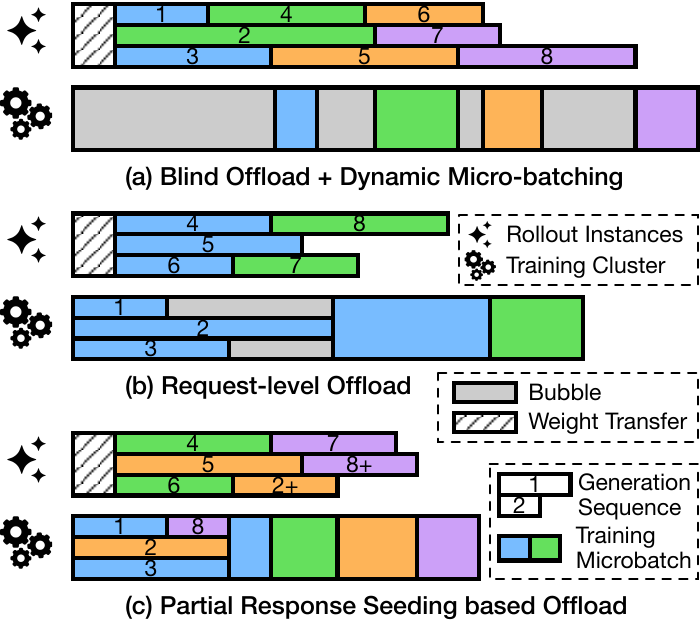}
\caption{\sys minimizes \trainers idling with an adaptive partial response seeding mechanism.}
\label{fig:response_seeding}
\end{figure}

\begin{algorithm}[t]
\SetAlgoLined
\SetAlgoNoEnd
\DontPrintSemicolon
\SetKwInOut{Input}{Input}
\SetKwInOut{Output}{Output}
\newcommand\myCommentStyle[1]{\small\textcolor{gray}{#1}}
\SetCommentSty{myCommentStyle}
\SetKwComment{Comment}{// }{}
\Input{$N_{\text{resv}}$: Number of local rollout engines in the reserved \trainers; $\eta$: Adaptation rate; $T_{\text{init}}$: Initial seeding window, $S$: Total number of training steps.}
$\mathcal{M} \leftarrow \emptyset$ \Comment{scheduler memory}
$T_{\text{seed}} \leftarrow T_{\text{init}}$ \Comment{rollout time window on the \trainers}
$N_{\text{prem}} \leftarrow N_{\text{resv}} $ \Comment{initialize max amount of preemptible instances the same as reserved rollout engines}
\For{$s \leftarrow 1$ \textup{\textbf{to}} $S$}{
 \textsc{ExecuteStep}$\left(T_{\text{seed}}, N_{\text{prem}}\right)$\;
 $\overline{n}_{\text{prem}}, \hat{n}_{\text{prem}} \leftarrow \textsc{MonitorInstAvail}\left(\right)$\; \label{eq:execute_rl_step}
$t^{\text{wait}}_{\text{train}}, T^{\text{wait}}_{\text{remote}} \leftarrow \textsc{GetIdleTime}\left(\right)$\; \label{eq:record_wait_time}
 $t_{\text{train}}, t_{\text{remote}} \leftarrow \textsc{GetComputeTime}\left(\right)$\; \label{eq:record_total_time}
 \Comment{update schedule}
 $T_{\text{seed}} \leftarrow T_{\text{seed}} +\frac{t^{\text{wait}}_{\text{train}}-t^{\text{wait}}_{\text{remote}}}{\eta}$\; \label{eq:t_seed_adjust}
 $N_{\text{prem}} \leftarrow \frac{t_{\text{remote}}\overline{n}_{\text{prem}}+T_{\text{seed}}N_{\text{resv}}}{t_{\text{train}}}$\; \label{eq:max_instances_adjust}
 \If{$\overline{n}_{\text{prem}} = \hat{n}_{\text{prem}}$}{
    \Comment{memorize schedule optimized under $\hat{n}_{\text{prem}}$}
    $\mathcal{M}\left[\hat{n}_{\text{prem}}\right] \leftarrow T_{\text{seed}}$
    \label{eq:update_memory}
 }
 \If{$\hat{n}_{\text{prem}}\in\mathcal{M}$}{
    \Comment{retrieve latest schedule optimized for $\hat{n}_{\text{prem}}$}
    $T_{\text{seed}} \leftarrow\mathcal{M}[\hat{n}_{\text{prem}}]$ \label{eq:retrieve_from_memory}
 }
}
\caption{Adaptive partial response seeding.}
\label{algo:adaptive_seeding}
\end{algorithm}

Because \sys offloads rollout from the training cluster to a separate pool of preemptible instances, it faces the same resource idleness issue as shown in \autoref{fig:arch_compare}(a), attributed to the dependency between rollout and training. To overlap the execution of the training cluster and remote rollout instances, we can employ dynamic microbatch pipelining, similar to \cite{zhong2025streamrl}. The training cluster collects responses from the rollout manager as soon as they are generated, until a minimum microbatch size of $m_b$ is reached, then immediately begins training of the microbatch, as illustrated in \autoref{fig:response_seeding}(a). If more than $m_b$ responses arrive at once, they are gathered in a single microbatch. Since gradients are accumulated across all responses, they can be collected and batched without preserving the original order in which prompts are issued to rollout instances. Notably, even in the co-located architecture, training is already executed in a series of microbatches, because all generated responses cannot fit into a single training batch constrained by GPU memory. Hence, dynamic micro-batching does not hurt compute efficiency.

If we blindly offload all rollout computation, even with dynamic micro-batching, the training cluster still suffers from significant bubbles, especially when insufficient remote rollout instances are available. Specifically, the training cluster must wait for rollout instances to receive model weights at the beginning of each step, and wait between microbatches for responses to be generated.

To balance the execution between the training cluster and remote rollout instances, \sys must dynamically adjust the offloaded rollout workload \textit{to adapt to preemptible resource availability}. A straightforward approach is to assign a specified number of rollout requests for the training cluster to generate locally while offloading the rest to remote instances, as is shown in \autoref{fig:response_seeding}(b). However, this offloading strategy is too coarse-grained. Since response lengths are highly unpredictable, the training cluster may be stuck generating long-tail responses even after receiving sufficient responses for training.

Our insight is that instead of controlling the number of rollout samples to offload, we can bound the trainer's rollout work in \textit{time} to make progress predictable. To this end, we design a partial response seeding mechanism: \sys allows \trainers to rollout only within a fixed time window at the beginning of each step, and then it transitions to training. For long-tail responses, the \trainers "seeds" a part of the response for \rollouters to continue from, as illustrated in \autoref{fig:response_seeding}(c) for response 2. Since \rollouters only need to compute a single prefill over the already generated tokens, migrating partially generated responses introduces minimal overhead (see \autoref{sec:design_request_migration}).

However, determining the optimal seeding duration remains non-trivial. If set too long, training is unnecessarily delayed; if too short, \trainers still experience bubbles waiting for responses. Moreover, the optimal setting is dynamic due to two key factors. In addition to the fluctuating number of available preemptible instances for rollout, the average response length tends to grow as RL training progresses~\cite{yu2025dapo}. These factors cause unpredictable changes in rollout and training times throughout the RL training process.

Beyond the challenge of identifying the optimal seeding window, another question is how many preemptible instances should we actually use, even when availability is unlimited. Given their cost advantages, we can follow the established practices in \cite{thorpe2023bamboo,miao2024spotserve,duan2024parcae} and use as many instances as available to maximize the generation speedup. However, the training stage still imposes a lower bound on step time. Hence, we must avoid over-provisioning remote \rollouters.

We present an adaptive scheduling algorithm in \autoref{algo:adaptive_seeding} that addresses both initial idling on the \trainers and resource waste of preemptible instances. 
Each remote rollout instance uses the same number of GPUs as one local rollout engine’s tensor parallel size.
The algorithm dynamically adjusts the seeding window $T_{\text{seed}}$ and enforces a maximum number of allowed remote \rollouters $N_{\text{prem}}$ by monitoring step time statistics. In each step, \sys tracks idle time on both the \trainers $t^{\text{wait}}_{\text{train}}$ and remote \rollouters $t^{\text{wait}}_{\text{remote}}$. $t^{\text{wait}}_{\text{train}}$ represents the idle time on the \trainers, waiting for sufficient responses to fill a microbatch.
$t^{\text{wait}}_{\text{remote}}$ measures how long remote instances wait for the \trainers to complete the current step, after they generate the last response. Ideally, to minimize the total step time, we should minimize $T_{\text{seed}}+t^{\text{wait}}_{\text{train}}$, as the \trainers's completion marks the step completion.
However, due to the unpredictable nature of responses arrivals, generation lengths, and instance availability, $t^{\text{wait}}_{\text{train}}$ and $t^{\text{wait}}_{\text{remote}}$ are highly indeterministic. They are also intertwined and are both correlated with $T_\text{seed}$. Hence, \sys employs a feedback-driven mechanism to incrementally tune $T_{\text{seed}}$, maintaining stability across steps under fluctuations while adapting to evolving workload patterns.
\sys should increase $T_{\text{seed}}$ when observing a significant $t^{\text{wait}}_{\text{train}}$. Yet, $T_{\text{seed}}$ cannot grow indefinitely as it would delay the overall step time, which will be reflected in a longer $t^{\text{wait}}_{\text{remote}}$. As shown in line~\ref{eq:t_seed_adjust} of \autoref{algo:adaptive_seeding}, \sys adjusts $T_{\text{seed}}$ by balancing between the two objectives, with a scale factor $\eta$ applied to the adjustment delta. 

The tuning in line~\ref{eq:t_seed_adjust} needs gradual progression to converge after the number of remote instances changes. To mitigate the re-tuning overhead when many instances join or are preempted during a step, \sys employs a memorization mechanism in line~\ref{eq:retrieve_from_memory} to directly start from the latest $T_{\text{seed}}$ optimized under $\hat{n}_{\text{prem}}$ instances, where $\hat{n}_{\text{prem}}$ is the number of active \rollouters available before the start of the subsequent step.  The scheduler memory $\mathcal{M}$ is continuously updated after each step in line~\ref{eq:update_memory}, provided no instance changes occurred during the step, i.e., only when $\hat{n}_{\text{prem}} = \overline{n}_{\text{prem}}$. $\overline{n}_{\text{prem}}$ is the number of instances averaged over the duration of a step.

To prevent \sys from over-allocating remote rollout instances that would yield no further performance improvement, in line~\ref{eq:max_instances_adjust}, \sys sets the upper bound $N_{\text{prem}}$ by computing how many instances are required for the rollout stage to take less time than $t_{\text{train}}$. $t_{\text{train}}$ is the effective time the \trainers spent on training in a step, i.e., excluding idle periods. To preclude the impacts of $T_{\text{seed}}$ , we assume rollout is solely processed by remote instances when computing $N_{\text{prem}}$, where $t_{\text{remote}}\overline{n}_{\text{prem}} + T_{\text{seed}} n_{\text{resv}}$ is the total rollout workload. $n_{\text{resv}}$ is the number of rollout engines (instances) the \trainers is divided into during rollout seeding. The rollout manager in \sys keeps tracks of instance availability and allocates new instances upon availability. If there are already $N_{\text{prem}}$ remote instances, \sys will not allocate a new instance even if more are available.

With adaptive rollout offload minimizing \trainers idle time, \sys  falls back to colocated rollout when spot capacity is unavailable or frequently preempted, without stalling progress.
Next, we explore how \sys enables no-waste preemption handling and continuous load balancing with token-level response collection.

\subsection{Live Request Tracking and Migration}
\label{sec:design_request_migration}

\begin{figure}
\centering
\includegraphics[width=0.95\linewidth]{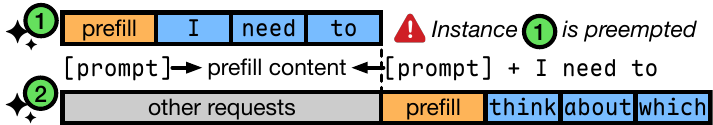}
\caption{\sys collects responses at token granularity and migrates requests upon instance preemption, incurring only the cost of an additional prefill.}
\label{fig:request_migrate}
\end{figure}

\begin{algorithm}[t]
\SetAlgoLined
\SetAlgoNoEnd
\DontPrintSemicolon
\SetKwInOut{Input}{Input}
\SetKwInOut{Output}{Output}
\newcommand\myCommentStyle[1]{\small\textcolor{gray}{#1}}
\SetCommentSty{myCommentStyle}
\SetKwComment{Comment}{// }{}
\SetKwProg{Fn}{function}{}{end}
\SetKwProg{Proc}{procedure}{}{end}
\Input{$\mathcal{I}$: Set of rollout instances; $\mathcal{P}$: Inference batching profile table; $\Theta$: Maximum pending requests threshold.}
\Fn{\textsc{SelectInstance}$\left(\mathcal{I}\right)$}{
\While{true \label{eq:select_instance_while_loop}}{
    $\mathcal{C} \leftarrow \emptyset$\;
    \ForEach{$i \in \mathcal{I}$}{
        $m^{\text{pending}}_{i} \leftarrow \textsc{QueryPending}\left(i\right)$\; \label{eq:select_instance_query_instance_pending}
        \If{$m^{\text{pending}}_i < \Theta$}{
            $\mathcal{C}\leftarrow \mathcal{C} \cup \{i\}$\;
            \label{eq:add_to_candidate}
        }
    }
    \If{$\mathcal{C} \neq \emptyset$}{
        $i\leftarrow \argmin_{i\in \mathcal{I}} m^{\text{pending}}_i$\; \label{eq:select_shortest_queue}
        \Return{$i$}\; 
    }
    \Else{
        \textsc{WaitAnyCompletion}$\left(\right)$\; \label{eq:wait_any_complete}
    }
}
}
\Proc{\textsc{ContinuousLB}$\left(\mathcal{I},\mathcal{P}\right)$}{
\While{true}{
    \ForEach{$i \in \mathcal{I}$}{
        $m^{\text{pending}}_{i}\leftarrow \textsc{QueryPending}(i)$\; \label{eq:lb_query_pending}
        $m^{\text{exec}}_{i}\leftarrow \textsc{QueryExecuting}(i)$\;
        \label{eq:lb_query_executing}
    }
    \If{$\exists i, m^{\text{pending}}_{i} =0$ and $\exists k,m^{\text{pending}}_{k}>0$ \label{eq:lb_check_pending}}{
        $j\leftarrow\argmax_{k\in \mathcal{I}} m^{\text{pending}}_{k}$\;
        \label{eq:lb_find_most_load_pending}
        \Comment{migrate a single request}
        $\textsc{MigrateReqs}\left(j\rightarrow i, 1\right)$\; \label{eq:lb_migrate_req_pending}
    }
    \ElseIf{$\exists i, m^{\text{exec}}_{i}=0$ \label{eq:lb_check_exec}}{
        $j\leftarrow \argmax_{k\in\mathcal{I}} m^{\text{exec}}_{k}$\;
        \label{eq:lb_find_most_load_exec}
        $B\leftarrow \textsc{GetBatchingPlateu}\left(\mathcal{P}\right)$\;
        \label{eq:lb_compute_batch_size}
        $r\leftarrow \max (m^{\text{exec}}_{j}- B,0)$\;
        \label{eq:lb_compute_exec_req_to_migrate}
        \Comment{migrate $r$ requests}
        $\textsc{MigrateReqs}\left(j\rightarrow i, r\right)$\;
        \label{eq:lb_migrate_exec_reqs}
    }
}
}
\caption{\sys's load balancer.}
\label{algo:load_balancer}
\end{algorithm}

Since a rollout instance can be preempted at any instant, requests routed to it may not complete generation when preemption occurs. To preserve the correctness of the RL workflow, we cannot drop any of preempted requests. However, simply retrying the request on another rollout instance from the original prompt would result in significant progress loss, particularly when most tokens of a sample has been generated. To minimize lost progress and redundant computation upon a preemption, our insight is to collect the response at token granularity. For each request, the rollout manager spawns an asynchronous task to track and receive the response tokens in a streaming manner. When an instance is preempted, \sys still preserves partially generated responses for requests routed to the instance. For each partially generated sample, \sys migrates the request to one of the healthy instances to continue generation  (\autoref{fig:request_migrate}). Similar to \autoref{sec:design_response_seeding}, the redirected instance only performs a prefill operation on the concatenated prompt and previously generated tokens, incurring negligible overhead compared to generating from the beginning (original prompt).

\subsubsection{Continuous Load Balancing}
Such token-level response collection not only reduces the costs of a preemption, it also empowers \sys with the ability to flexibly migrate and redistribute samples across instances, allowing continuous load monitoring and balancing.

We present the load balancer logic for \sys in \autoref{algo:load_balancer}.
It is composed of two main components: \textsc{SelectInstance} is used for initial candidate instance selection when a generation request is first scheduled, and re-routing when the previously selected instance is preempted. \textsc{ContinuousLB} is a background monitor task to continuously migrate requests from overloaded instances to underloaded ones as needed.

\textsc{SelectInstance} endorses the classical join the shortest queue (JSQ) scheduling policy widely used in web servers. It routes the generation request to the instance with the minimum number of pending requests (line~\ref{eq:select_shortest_queue}), i.e., requests that are already sent to the instance but have not been scheduled to execute yet. In the traditional JSQ policy, a request is immediately dispatched to an instance upon receiving it. Such a strategy works well in typical web servers of CPU-based processing, where requests are mostly homogeneous in the way that they take roughly the same amount of time to process. However, in LLM generation, instances with the most pending requests could complete the earliest due to variance in generation lengths. If all requests are immediately dispatched, we may need to frequently migrate requests to balance the load, causing unnecessary overhead. Instead, \sys adopts a \textit{delayed dispatch} approach, where we limit the number of outstanding pending requests to $\Theta$ for each instance. If all instances are already occupied with more than $\Theta$ pending requests, \sys waits for any of the in-flight request to finish (line~\ref{eq:wait_any_complete}) and rechecks the pending status (line~\ref{eq:select_instance_while_loop}), holding the request until one of the instances becomes available.

Once all requests are dispatched, \sys monitors and dynamically rebalances load with \textsc{ContinuousLB}. In lines~\ref{eq:lb_query_pending}--\ref{eq:lb_query_executing}, \sys tracks both the number of pending requests $m^{\text{pending}}_i$ and the number of currently executing requests $m^{\text{exec}}_{i}$ for each instance.  \sys first checks if any instance $i$ has no pending requests while other instances have (line~\ref{eq:lb_check_pending}). \sys migrates pending requests from the most overloaded instance $j$ to $i$, one request at a time (line~\ref{eq:lb_migrate_req_pending}). If instance $i$ has enough capacity, the migrated request will be immediately scheduled. In this case, \sys keeps migrating more requests to instance $i$ until it is saturated, i.e., subsequent requests to $i$ will queue up.

If there are no pending requests on all instances, \sys then checks if any instance $i$ is completely idle (line~\ref{eq:lb_check_exec}), i.e., is not executing any request. In this scenario, \sys finds the most loaded instance $j$ with the largest $m^{\text{exec}}_{j}$ (line~\ref{eq:lb_find_most_load_exec}).  Different from the scenario with pending requests, migrating executing requests may not lead to earlier completion due to the batching effects of LLMs. If $m^{\text{exec}}_{j}$ is small enough, the generation is completely memory-bound, removing requests from $j$ leads to no improvement in inter-token latency (ITL), but instead a linear decrease in generation throughput. However, if $m^{\text{exec}}_{j}$ is beyond the point where further increases in batch size yield only marginal throughput gains, migrating a part of the requests out of $j$ helps speed-up the overall generation. In line~\ref{eq:lb_compute_exec_req_to_migrate}, \sys determines the number of requests $r$ to migrate from $j$ to $i$ by clamping $m^{\text{exec}}_{j}$ to the batch size $B$ where the generation throughput plateaus, where $B$ is computed from a profile table $\mathcal{P}$ of throughput under different batch sizes (line~\ref{eq:lb_compute_batch_size}). Instead of offline profiling, $\mathcal{P}$ is \textit{online captured} by \sys during the previous step's rollout, and is \textit{continuously calibrated} to account for the current average context length. We also tried directly incorporating both batch size and the context length into $\mathcal{P}$,  but found it difficult to fit the performance model across two dimensions, resulting in worse estimates. We note that since $\mathcal{P}$ is only established after the first step, \textsc{ContinuousLB} begins to migrate executing requests from the second step onward. 

At this point, through adaptive rollout offload and migration-based load balancing, \sys can maximize effective compute on the \trainers and remote \rollouters, while efficiently handling preemptions. Next, we discuss how \sys decouples the weight transfer logic from the training and generation workers.

\subsection{Pull-based Weight Transfer} \label{sec:weight_transfer}

\begin{figure}
\centering
\includegraphics[width=0.95\linewidth]{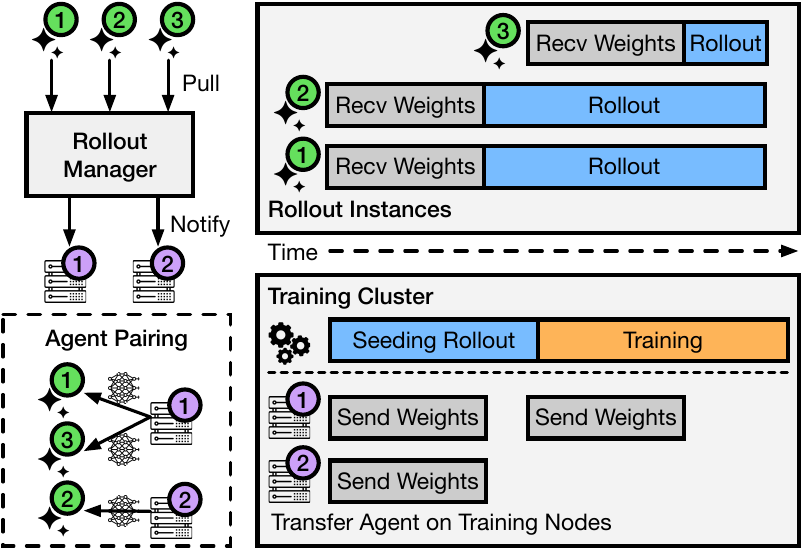}
\caption{Pull-based weight transfer enables newly allocated rollout instances to be quickly provisioned with the latest model weights without blocking existing workers.}
\label{fig:weight_transfer}
\end{figure}

After the training stage and the model is updated, \sys will reshard model weights for seeding rollout on \trainers, in the same way as co-located RL frameworks. The resharding within the \trainers is carried out over fast interconnects like NVLink and RDMA, which can be significantly faster than the bandwidth between \trainers and \rollouters. In modern GPU clusters, the frontend and backend networks are typically separated, with the high-capacity backend network dedicated for GPU data traffic within the cluster~\cite{gcp_network,aws_network}. On public clouds, even if \trainers and \rollouters are located in the same datacenter, they can be limited by the slower frontend network~\cite{gcp_network}. 

Besides the asymmetric network bandwidth problem, if we use the synchronized weight update approach in co-located frameworks that transfers weights only after each step, an instance joined midway through a step cannot process requests until the next step. Also, the completion of weight update can be blocked by \rollouters with poor network bandwidth.

To unblock the \trainers for rollout seeding and to immediately transfer the latest weights to a rollout instance once they are allocated, \sys employs a pull-based transfer agent to asynchronously transfer weights, as shown in~\autoref{fig:weight_transfer}. 
The transfer agent is a separate process residing on each training node and rollout instance. During the intra-cluster all-gather, each training node copies the full model weights from GPU to a pre-allocated CPU buffer managed by the transfer agent. After that, the \trainers immediately starts seeding rollout, instead of waiting for the weight delivery to all \rollouters. Each \rollouter is paired with a weight transfer agent in a round-robin way and establishes a peer-to-peer connection. On initial registration or model update, a \rollouter will independently pull the latest weight and start generation once the transfer finishes, without affecting other \rollouters and \trainers.

%% file: implementation.tex
\section{Implementation}

We implement \sys based on a derived version of veRL~\cite{sheng2025hybridflow} in 2.7K lines of Python and 1.7K lines of Rust. \sys supports PyTorch FSDP~\cite{zhao2023fsdp} and Megatron~\cite{shoeybi2019megatron} for training and uses SGLang~\cite{zheng2024sglang} for rollout. 

\noindent \textbf{Rollout manager.}
We implement the rollout manager as a RESTful API web service using Rust's asynchronous framework with Tokio~\cite{tokio} and Axum~\cite{axum}. The manager monitors instance availability and allocates new rollout instances when permitted, ensuring the total count does not exceed the upper bound $N_{\text{prem}}$. It keeps track of idle waiting time and effective compute time reported by the rollout instances and the training cluster, which are used to compute $T_{\text{seed}}$ and configure the training cluster for the next step. The rollout manager also periodically probes each rollout instance's $m^{\text{pending}}_i$ and $m^{\text{exec}}_i$ for load balancing. For each rollout request, an asynchronous task is launched to track the request's entire lifetime, collecting response tokens as they are generated. If the routed instance is preempted, the tracking task would detect a connection-closed error, it then immediately requests another healthy instance from the load balancer and migrates the request. 

When a new rollout instance is allocated, it registers with the manager, which then assigns it to a weight sender agent, as described in \autoref{sec:weight_transfer}. The manager maintains the weight version of each instance and notifies the paired sender agent to transfer weights if an instance is not up-to-date. The manager only routes requests to instances that have loaded the latest weights.

\noindent \textbf{Trainer workers.}
Using components from veRL~\cite{sheng2025hybridflow}, we implement the trainer worker to support dynamic micro-batching as discussed in \autoref{sec:design_response_seeding}, with an asynchronous task running on the CPU collects complete responses from the rollout manager and packs them into micro-batches.

\noindent \textbf{Transfer agents.}
We assume weight transfer is conducted over the frontend network (\autoref{sec:weight_transfer}) and thus does not contend with training or inference traffic on the backend network. Since RDMA networks between rollout instances and the training cluster may not be available, particularly in public cloud environments, we implement a TCP-based weight transfer engine in our prototype. To fully utilize the bandwidth of all available frontend NICs, \sys uses multiple I/O threads with each handling a different weight shard, while each sender agent transfers weights to multiple rollout instances simultaneously. 
The design of \sys' pull-based weight transfer is agnostic to the underlying transport. When RDMA interconnects are available, we can easily integrate RDMA optimized transport engines (e.g., Mooncake~\cite{qin2025mooncake} and NIXL~\cite{nixl}) into \sys.

%% file: evaluation.tex
\section{Evaluation}
\label{sec:eval}
In this section, we evaluate \sys against co-located and disaggregated RL frameworks with models from 8B to 32B, comparing both performance and cost efficiency. We then breakdown the benefits of different components of \sys.

\begin{table}[t]
\centering
\caption{Specification of public cloud instances. The cost estimation details are in \autoref{sec:cloud_instance_cost}.}
\label{tab:cluster-setup}
\small
    {
    \begin{tabular}{lcc}
        \toprule
         & On-Demand Training & Spot Rollout \\ 
        \midrule
        vCPUs & 208 & 52 \\
        \midrule
        GPU & 8xH100 & 2xH100 \\
        \midrule
        Network & (4+1)x200~Gbps & 50 Gbps \\
        \midrule
        Cost & \$83.79/h & \$5.32/h \\
        \bottomrule
    \end{tabular}
    }
\end{table}

\begin{figure}[t]
\centering
\includegraphics[width=0.98\linewidth]{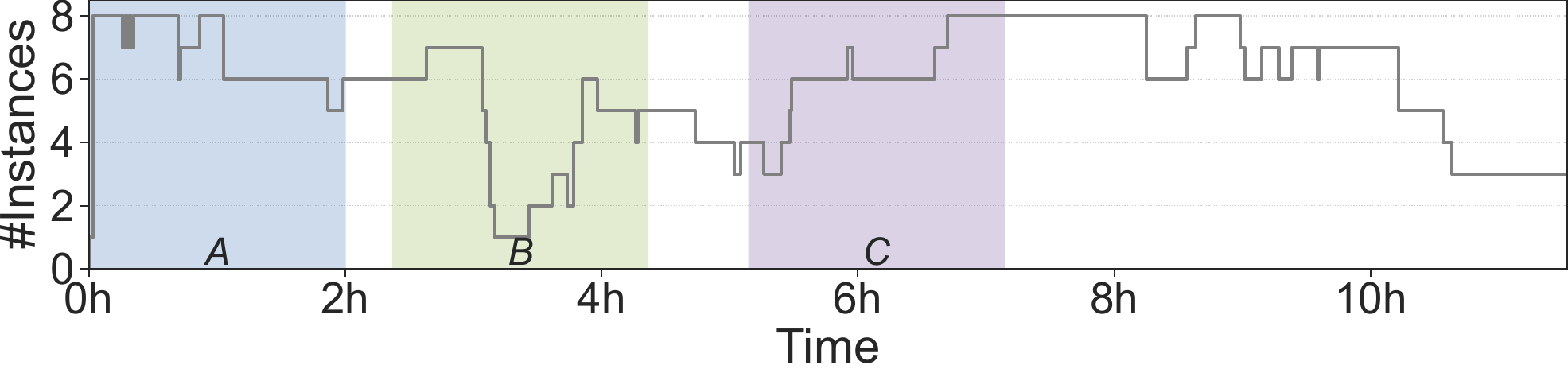}
\caption{The complete 12-hours availability trace for \texttt{2xH100} instances and the three 2-hours segments (A, B, C) extracted.}
\label{fig:overall-trace}
\end{figure}

\subsection{Setups}

\begin{figure*}[t]
\centering
\includegraphics[width=0.99\linewidth]{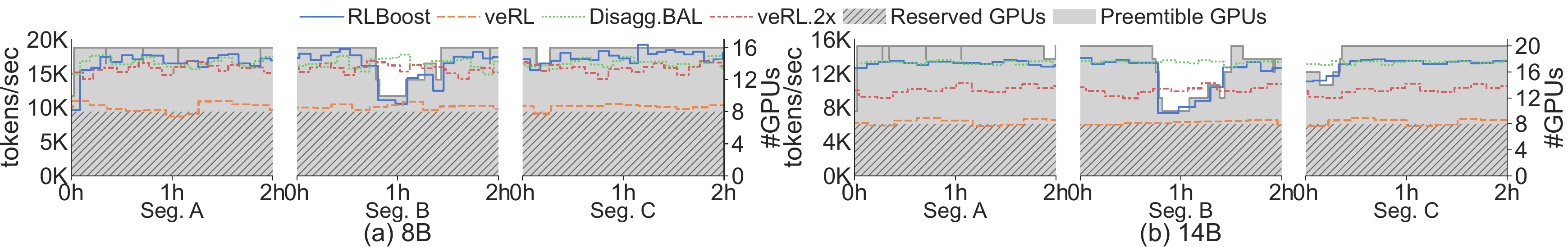}
\caption{\textbf{[Overall evaluation]:} Throughput over each trace segment for Qwen-8B and Qwen-14B. The number of reserved GPUs and the number of preemptible GPUs allocated and used by \sys is also shown. veRL, veRL.2x and \disaggbl only use reserved GPUs, with veRL.2x and \disaggbl using more reserved GPUs than \sys.}
\label{fig:overall-eval-timeline-part1}
\end{figure*}

\begin{figure}[t]
\centering
\includegraphics[width=0.99\linewidth]{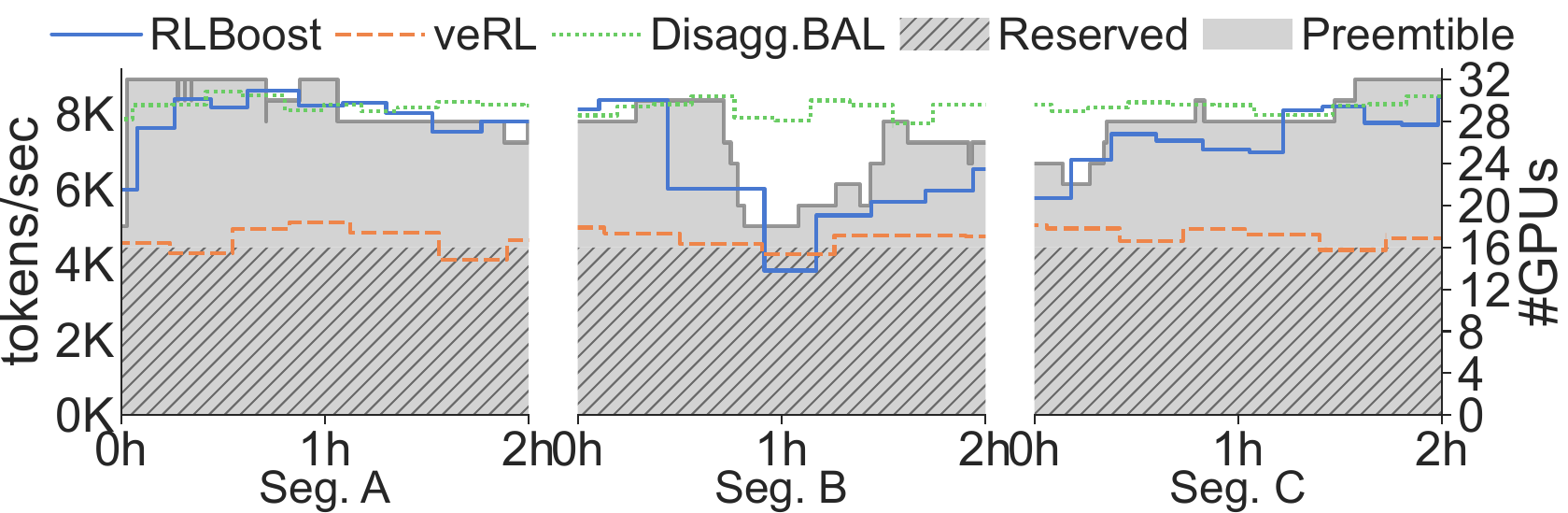}
\caption{\textbf{[Overall evaluation]:} Throughput over each trace segment for Qwen-32B.}
\label{fig:overall-eval-timeline-part2}
\end{figure}

\begin{figure*}[t]
\centering
\includegraphics[width=0.99\linewidth]{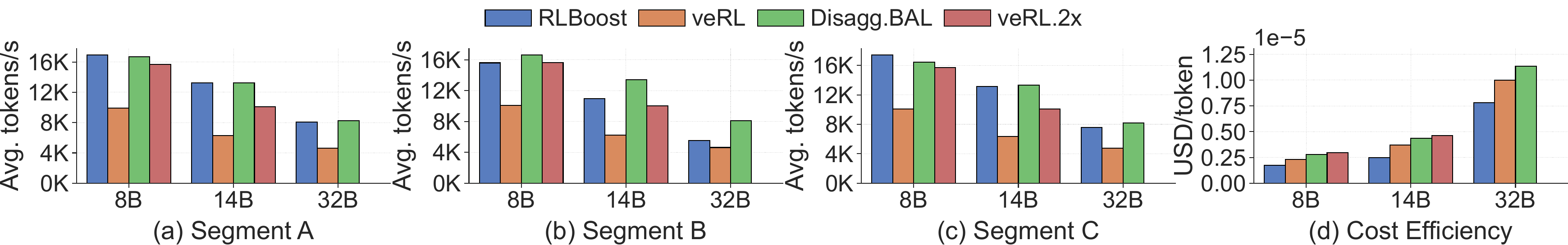}
\caption{\textbf{[Overall evaluation]:} Average throughput and cost efficiency across all three trace segments.}
\label{fig:overall-eval-bar}
\end{figure*}

\noindent
\textbf{Hardware settings.} We evaluate \sys using H100 GPU instances on a public cloud. For the training cluster, we target on-demand (reserved) instances each fully equipped with 8~H100 GPUs. For preemptible rollout instances, we target spot instances each equipped with 2~H100 GPUs, since fragmented instances generally have better availability~\cite{duan2024parcae}. We list the detail specifications of both instance types in \autoref{tab:cluster-setup}. Notably, the \texttt{8xH100} instances feature 4 backend NICs with an RDMA featured networking stack, whereas the \texttt{2xH100} instances are limited to a single frontend vNIC. Consequently, \texttt{8xH100} must rely on its single 200~Gbps frontend NIC to communicate with \texttt{2xH100}. All GPUs in a single instance are fully connected with 900~GB/s NVLink. For better representativeness, in \autoref{tab:cluster-setup}, we calculate the average cost of on-demand and spot instances with the same GPUs across different cloud providers in different regions. These per-instance costs are used to compute cost efficiency of compared systems.

\noindent
\textbf{Workloads.} We utilize models from the popular Qwen~3 family~\cite{qwen3technicalreport} as base models, spanning from 8B to 32B.  We provide their configurations in~\autoref{sec:model_config}.
We employ the synchronized (on-policy) GRPO algorithm, the current mainstream RL algorithm for LLMs. We note that many newly proposed algorithms, e.g., DAPO~\cite{yu2025dapo} and GMPO~\cite{zhao2025gmpo} are derived from GRPO and share similar workload patterns. 

We use a math dataset OpenR1-Math~\cite{openr1math-220k} to train the models, which has a maximum response length of 14K tokens. Following the practice in \cite{sheng2025hybridflow,zhong2025streamrl}, we use a global batch size of 128 prompts, with a GRPO group size of 8. The models are trained with FSDP~\cite{zhao2023fsdp}. To demonstrate \sys across different cluster scales, we use a single on-demand \texttt{8xH100} instance for training 8B and 14B models, while for 32B models we use two \texttt{8xH100} instances. 

\noindent
\textbf{Traces.} For evaluation reproducibility, we follow \cite{duan2024parcae,miao2024spotserve,jang2023oobleck} to take real spot instances trace from~\cite{thorpe2023bamboo} and replay them on on-demand instances. Following the practice in~\cite{duan2024parcae,miao2024spotserve}, we extract three representative 2-hours segments in as the preemption traces for the \texttt{2xH100} instances, and describe their characteristics in ~\autoref{sec:spot_trace}. We note that all \texttt{8xH100} instances are reserved and will not be preempted.

\noindent
\textbf{Metrics.} Following~\cite{sheng2025hybridflow,fu2025areal}, we report the system performance in terms of effective training throughput. The throughput of a step is measured as the total number of tokens generated and trained in the step, divided by the time of the step.

\begin{figure}
\centering
\includegraphics[width=0.99\linewidth]{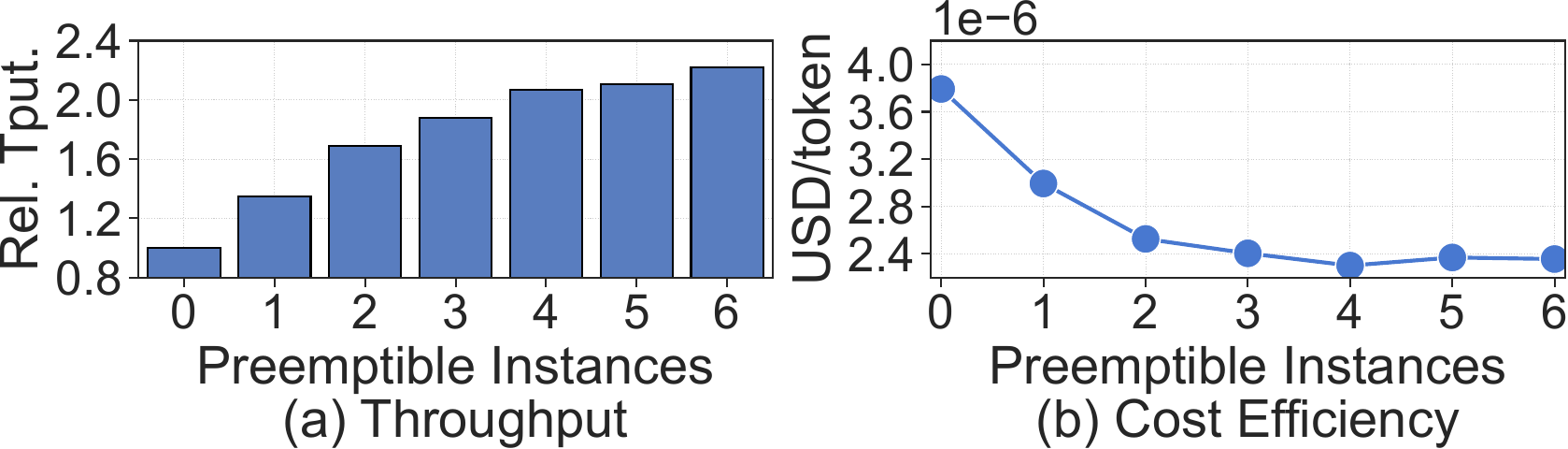}
\caption{\textbf{[Cost efficiency]:} Throughput and cost efficiency of \sys on Qwen-14B with a static number of preemptible rollout instances. 0 refers to only use the training cluster.}
\label{fig:ablation_cost_efficiency}
\end{figure}

\begin{figure}
\centering
\includegraphics[width=0.99\linewidth]{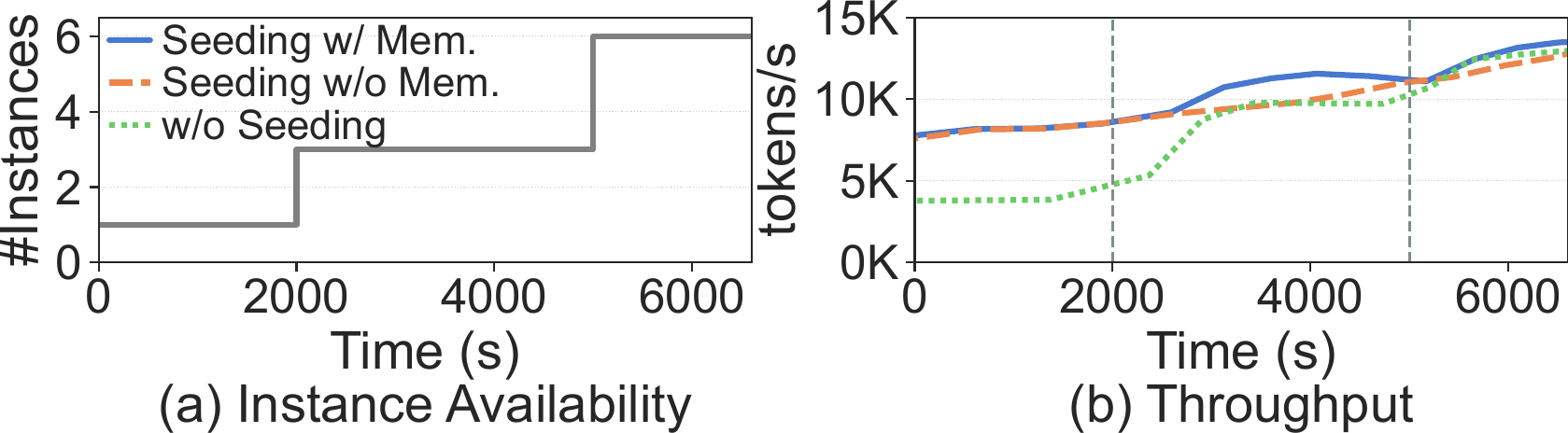}
\caption{\textbf{[Ablation study]:} The impacts of adaptive rollout offload with partial response seeding on Qwen3-14B.}
\label{fig:ablation_seeding}
\end{figure}

\subsection{Overall Evaluation}
\label{sec:eval_overall}
We compare the end-to-end performance of \sys with the following systems: 
\begin{itemize}[leftmargin=*]
    \item \textbf{veRL}~\cite{sheng2025hybridflow}: A state-of-the-art RL system under the co-located architecture. It features an optimized execution engine to efficiently manage and execute RL workflows for LLMs. We run veRL on the training cluster in each setup, i.e., a single \texttt{8xH100} instance for Qwen3-8B and Qwen3-14B, and two \texttt{8xH100} for Qwen3-32B. 
    
    \item \textbf{veRL.2x}: To evaluate how the cost efficiency varies for the co-located architecture when scaling up, we also run veRL with 2x more hardware resources, e.g., two \texttt{8xH100} instances for 8B and 14B models. veRL.2x is not evaluated on Qwen3-32B, as we do not have additional reserved \texttt{8xH100} instances.
    
    \item \textbf{Disagg.BAL} (Balanced): StreamRL~\cite{zhong2025streamrl} is a state-of-the-art disaggregated framework that targets asynchronous RL. However, it also includes several optimizations that can be applied to the on-policy setting. Since it is not open sourced, we implement a disaggregated framework using techniques from \cite{zhong2025streamrl}. In particular, it features a resource optimizer that determines the number of GPUs allocated to each stage that balances the workloads and process rollout results in microbatches to reduce bubbles. 
    We use it to calculate the optimal number of reserved \texttt{2xH100} instances for rollout, given the number of GPUs used in the training cluster. 
\end{itemize}

Note that none of these systems can take advantage of preemptible instances. We present the training throughput and \sys's GPU usage over the duration of each segment in \autoref{fig:overall-eval-timeline-part1} and \ref{fig:overall-eval-timeline-part2}. We see that \sys's throughput fluctuates as \texttt{2xH100} instance availability changes, while throughput of other compared methods remain relatively stable as they only use reserved GPUs. We observe that in segment A there are many tiny spikes in preemptible GPU usage, which are also shown in \autoref{fig:overall-trace}. These spikes occur at the timestamps where a running \texttt{2xH100} instance is preempted, but a new one can be immediately allocated (also observed in \cite{thorpe2023bamboo,mao2025skyserve}). \sys shows negligible throughput drops in these cases, demonstrating that  \sys can effectively handle request failures and quickly set up new instances as they are allocated. 

We note that the preemption patterns are different across 8B to 32B models. Limited by $N_{\text{prem}}$, \sys does not allocate all available instances, hence a preempted instance may not be in use. The throughput is computed at the end of each RL step. For Qwen3-32B, each step takes significantly longer time than 8B and 14B, therefore the throughput changes are not immediately reflected in the curves.  Nevertheless, the performance boost \sys brings with preemptible GPU resources tightly matches resource availability. 

We show the average throughput over each segment and training costs in \autoref{fig:overall-eval-bar}. Compared to veRL that only uses the reserved training cluster, \sys significantly increases throughput. Across three segments, \sys outperforms veRL by 1.66x, 1.97x and 1.51x for 8B, 14B and 32B models, respectively, in terms of average throughput. \sys even achieves up to 24\% higher throughput than veRL.2x, which uses two \texttt{8xH100}, since the FSDP training in veRL.2x spans across two nodes and suffer additional overheads.

The throughput of \sys matches Disagg.BAL when there are sufficient preemptible \texttt{2xH100} available. In this case, \sys offloads virtually all rollout computation to \texttt{2xH100} instances, and thus exhibits comparable performance to Disagg.BAL. For instance, \sys's average throughput is just 2\% higher than Disagg.BAL over segment A. However, Disagg.BAL completely disaggregates rollout from the training cluster, and thus is unable to dynamically adapt the training cluster's workload in response to preemptible instance availability. It also lacks efficient request migration and pull-based weight transfer mechanisms, preventing it from handling dynamic instance changes.

Across all segments, \sys improves the training cost efficiency by 34\%, 49\% and 28\% for 8B, 14B and 32B models, compared to veRL. In segment A where the overall instance availability is high, \sys's achieves higher cost savings compared to veRL, improving cost efficiency by 36\%, 53\% and 45\% across 8B-32B models. The cost efficiency in segment B is relatively poor due to frequent preemptions and low availability, where the improvement of \sys over veRL drops to 37\% even for 14B.

Since Disagg.BAL can only use reserved instances, it suffers from poor cost efficiency. Across all three segments, its per-token training cost is 62\%, 75\% and 45\% higher than \sys, for 8B, 14B and 32B models.

\subsection{Analysis of Cost Efficiency}
\subsubsection{Impact of Preemptible Instance Availability}\label{sec:ablation_avail}
In \autoref{fig:overall-eval-bar}(d), we show the cost efficiency of \sys under fluctuating instance availability. We now break down how the throughput and cost efficiency of \sys change under a steady setting with different numbers of preemptible rollout instances. We show the results in \autoref{fig:ablation_cost_efficiency}.

The throughput is relative to the case with no preemptible instances for rollout, where \sys falls back to the same workflow as veRL, in which rollout is fully executed on the training cluster.
Both the throughput and cost efficiency improves until they reach the saturation point where rollout is already accelerated enough to match the training speed of the training cluster. Even with only one instance, the throughput increases by 37\% and the per-token training cost reduces by 22\%.  With 6 instances,  throughput further increases by 64\% compared to a single instance, while the cost further reduces by 21\%. We note that the trend in cost efficiency depends on the relative resource (GPU) ratio between the training cluster and the rollout instance pool, rather than the absolute number of rollout instances. \sys efficiently scales to more rollout instances as the training cluster scales. Besides, our evaluation across different maximum response lengths in~\autoref{sec:ablation_len} shows that \sys can utilize more and more preemptible instances as length grows to improve throughput and cost efficiency. 

\subsubsection{Impact of Maximum Response Length}
\label{sec:ablation_len}

\begin{figure}
\centering
\includegraphics[width=0.99\linewidth]{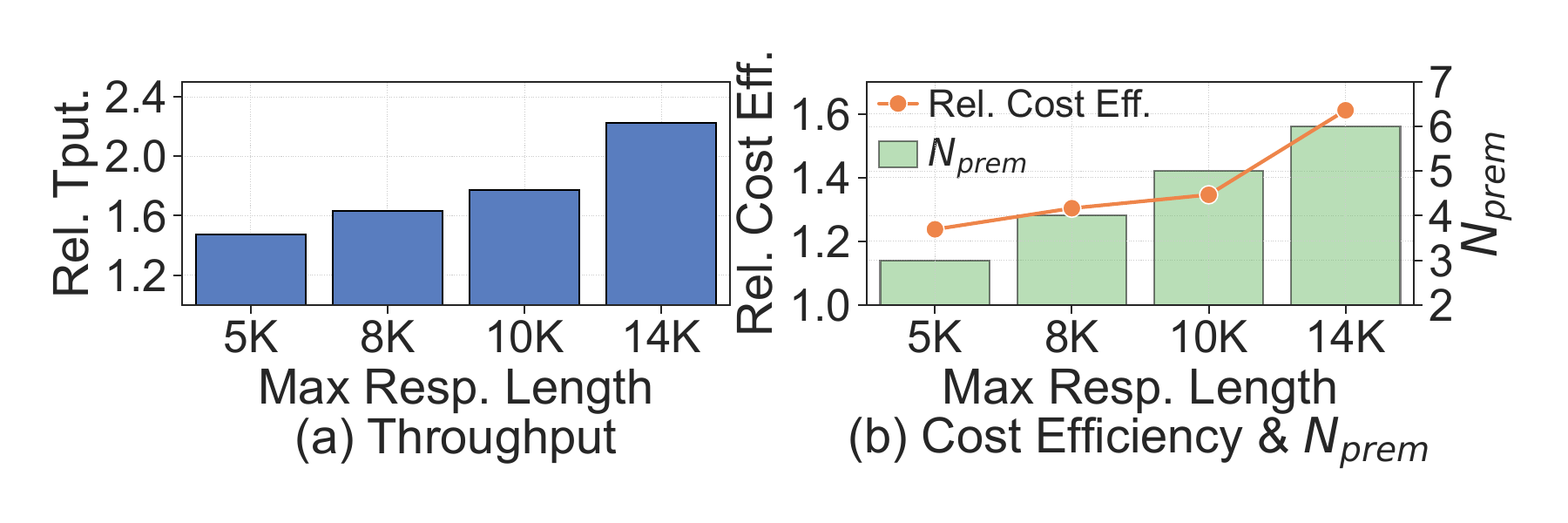}
\caption{\textbf{[Cost efficiency]:} Relative throughput and cost efficiency of \sys w.r.t. veRL on Qwen-14B using a single \texttt{8xH100} instance as the training cluster, under different max response length with corresponding optimal $N_{\text{prem}}$.}
\label{fig:ablation_len}
\end{figure}

We evaluate \sys under different maximum response lengths from 5K to 14K, and record the relative throughput and cost efficiency over veRL running on reserved instances in~\autoref{fig:ablation_len}. Due to the autoregressive computation pattern~\cite{kwon2023efficient} of LLM inference, rollout becomes more time-consuming than training as length grows. \sys automatically scales the number of preemptible instances to match the workload as in~\autoref{algo:adaptive_seeding}. As the optimal number of preemptible instances ($N_{\text{prem}}$) increases from 3 to 6, \sys boosts the relative throughput by 1.47x–2.22x and improves the relative cost efficiency by 1.24x–1.61x.

\subsection{Ablation Study}
\subsubsection{Impacts of Adaptive Rollout Offload}
We break down how \sys adapts the workloads on the training cluster to changes in preemptible instance availability. As described in \autoref{sec:design_response_seeding}, \sys adaptively offloads rollout to remote instances using a partial response seeding mechanism. In \autoref{fig:ablation_seeding}, we illustrate how \sys tunes $T_{\text{seed}}$ with \autoref{algo:adaptive_seeding} when instance availability changes. We construct a scenario where 5 out of 6 \texttt{2xH100} rollout instances are initially preempted, with substitute instances gradually becoming available after a period of time. Note that the initial preemptions are not displayed in \autoref{fig:ablation_seeding}(a).

We compare three solutions: response seeding using the complete \autoref{algo:adaptive_seeding} (full solution), a variant without scheduler memory, and a variant that disables response seeding. We observe that without seeding, \sys has to blindly offload all rollout requests to remote instances. Consequently, the training throughput is significantly lower during the initial stage, when only a single instance remains after preemptions. However, with more instances added, remote rollout is fast enough that $T_{\text{seed}}$ becomes negligible. Hence, \textit{w/o seeding} matches the performance of the full solution after all 6 instances become available. Over the duration of \autoref{fig:ablation_seeding}, \textit{w/o seeding} decreases the average throughput by 19\% compared with the full solution. Comparing seeding with and without memory, we find that the scheduler memory reduces the convergence time of $T_{\text{seed}}$ when new instances are allocated, resulting in a further 6\% average throughput increase.

\begin{figure}
\centering
\includegraphics[width=0.99\linewidth]{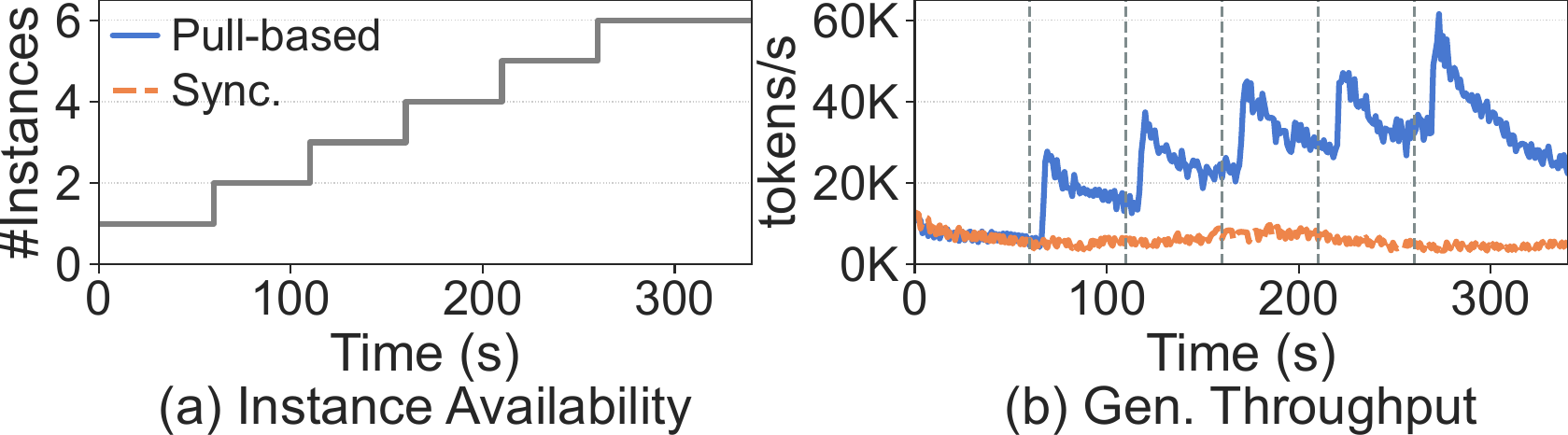}
\caption{\textbf{[Ablation study]:} Comparing pull-based and synchronized weight transfer as new instances are allocated within a step. We use Qwen3-14B.}
\label{fig:ablation_weigh_transfer}
\end{figure}

\begin{figure}
\centering
\includegraphics[width=0.95\linewidth]{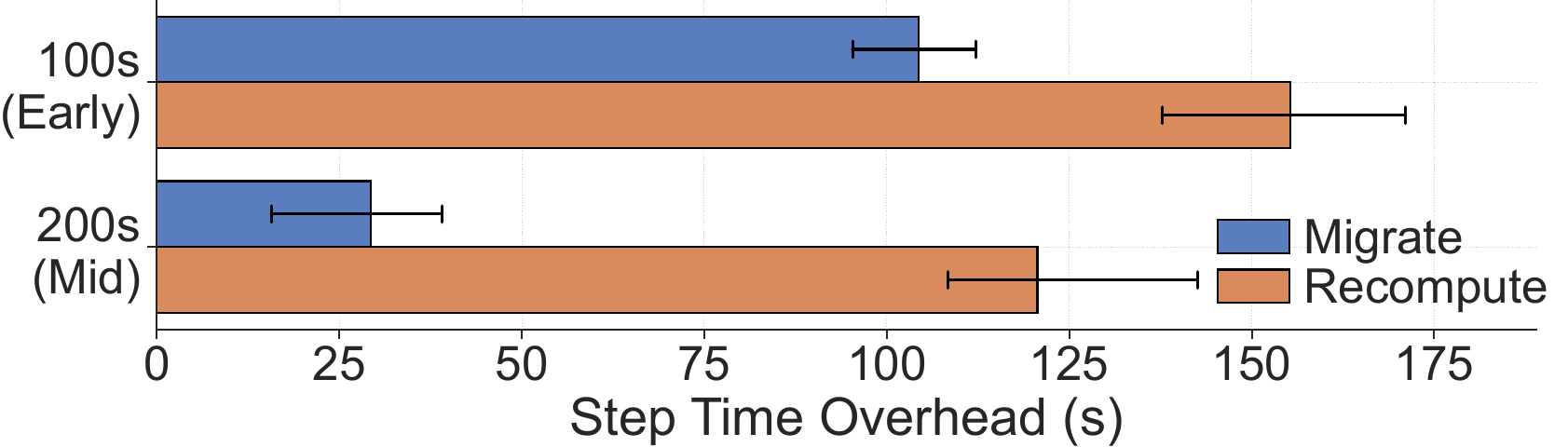}
\caption{\textbf{[Ablation study]:} Comparing different strategies for handling request failures upon preemptions on Qwen-14B. Error bars represent 95\% percentile intervals.}
\label{fig:ablation_fault_handling}
\end{figure}

\subsubsection{Impacts of Weight Transfer Paradigm}
The pull-based weight transfer agents in \sys decouple the transfer logic from training and rollout workers, allowing newly allocated preemptible instances to be quickly provisioned with the latest weights and participate in the current step's rollout. We construct a scenario where new instances are progressively allocated within a step. This represents the case that can be observed in \autoref{fig:overall-trace}: after simultaneous preemptions of many instances, new instances become available within a short period of time. 

We illustrate the effects in \autoref{fig:ablation_weigh_transfer}. Since we focus on the performance within a step, we cannot report effective training throughput, as it is computed per step. Instead, we report the total generation throughput aggregated over all rollout instances. The pull-based weight transfer enables \sys to immediately use a new instance for rollout, while the traditional synchronized weight transfer only makes use of new instances in the next step, causing substantial resource waste. We note that the generation throughput gradually drops after the initial surge when a new instance is added, resulting from the growing context length in the continuous batch as tokens are generated~\cite{kwon2023efficient}. We also demonstrate the impact of pull-based weight transfer agents when instances are preempted and immediately restart on another available node in~\autoref{sec:weight_transfer_restart}. 

\subsubsection{Impacts of Fault Handling}
To study how efficiently \sys handles request failures upon preemptions, we construct a scenario where 3 out of 6 rollout instances are preempted simultaneously at different points during the rollout of a step. We compare two fault handling strategies for requests routed to the preempted instances: our solution in \autoref{sec:design_request_migration} that collects responses at token granularity and migrates partially generated rollout requests (denoted as \textit{migrate}); and the traditional strategy that only collects complete responses, resulting in recomputing the entire request on a healthy instance (denoted as \textit{recompute}).

We analyze how different strategies impact step time. In \autoref{fig:ablation_fault_handling}, we report the step time overhead (increase) compared to the case with no preemption. We demonstrate two settings: preemptions at 100s (early point) and 200s (mid point) after the start of a step. For earlier preemptions at 100s, a limited number of tokens are generated for most requests, hence the cost of recomputation is not significant and migrate only reduces the overhead by 33\%. However, for preemptions at 200s, where many requests have already generated a large number of tokens, migrate reduces the overhead by 76\%.

\begin{figure}
\centering
\includegraphics[width=0.95\linewidth]{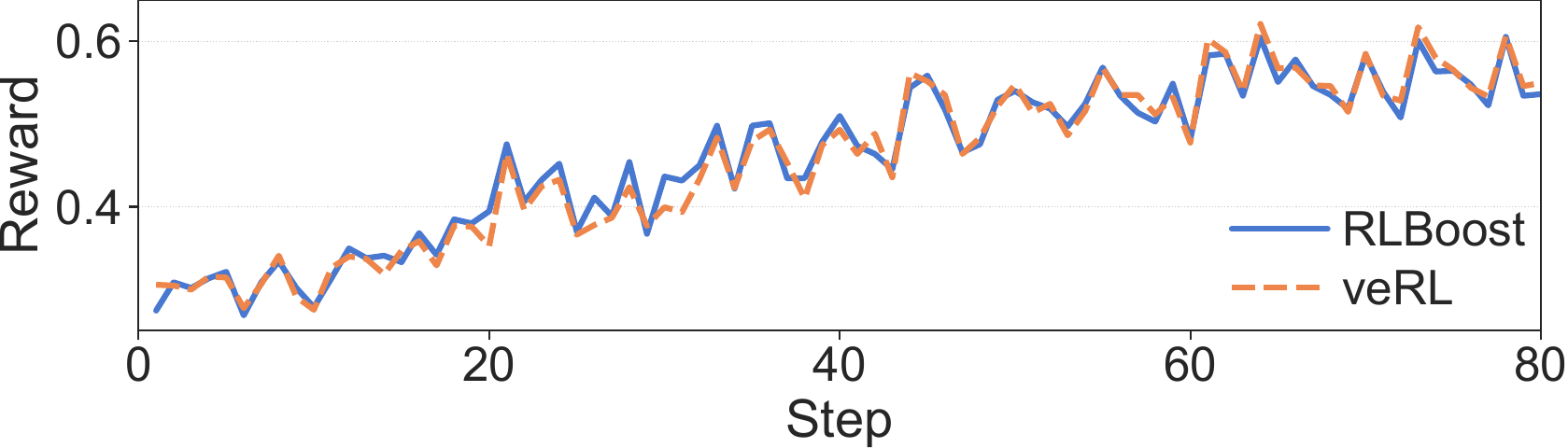}
\caption{\textbf{[Algorithm integrity]:} Rewards on Qwen-8B.}
\label{fig:ablation_loss_curve}
\end{figure}

\subsection{Algorithm Integrity}
Unlike many disaggregated frameworks~\cite{fu2025areal,zhong2025streamrl,han2025asyncflow,he2025history}, \sys maintains the well established synchronous RL algorithms. We compare the training reward curve of \sys with veRL in \autoref{fig:ablation_loss_curve}, where \sys is trained under the \texttt{2xH100} instance availability in \autoref{fig:overall-trace}. We present the curve for 8B since more steps can be trained within the same amount of time. The 14B and 32B models exhibit similar patterns.

As \sys makes no modifications to the synchronous GRPO algorithm and uses the same training settings, the reward curve of \sys closely matches that of veRL. We note that the reward eval at each step are not exactly the same. This is due to the temperature-based sampling in rollout, which is further complicated by the well-known nondeterministic behaviors in current LLM inference frameworks~\cite{he2025nondeterminism}.

%% file: discussion.tex
\vspace{-2mm}
\section{Discussion}

\paragraph{Supporting heterogeneous hardware.}

Since rollout instances operate independently from one another, \sys can exploit heterogeneous computing resources to further improve cost efficiency. Each rollout instance can be configured with different accelerators (varying GPU models or even TPUs) and parallelism strategies (different TP sizes). By leveraging real-time load and step time statistics in \autoref{algo:adaptive_seeding} and \ref{algo:load_balancer}, \sys can match each instance's computing capability, maximizing effective compute while ensuring load balance across heterogeneous instances. 

\paragraph{Weight transfer optimization.}
\sys currently implements a bipartite point-to-point weight transfer mechanism, where each rollout instance directly fetches weights from one of the training nodes. This strategy already fully utilizes all frontend NICs on training nodes within the same datacenter setting, where multiple network routes exist between rollout instances and the training cluster. 
Scaling to larger models would require more and larger GPU instances for both training and rollout. Hence, the per-instance and aggregated bandwidth increase accordingly. Consequently, weight transfer remains a small fraction of per-step time, which itself grows with model size. In addition, \sys can incorporate weight compression techniques~\cite{yao2025deltazip} to transfer compressed weight deltas between consecutive steps. 

\sys can be further optimized by building a dynamic broadcast tree~\cite{zhuang2021hoplite}, where only a subset of rollout instances retrieves weights from the training cluster while others receive them from peers. This optimization is beneficial when rollout instances are in a different datacenter, where cross-datacenter bandwidth is the bottleneck. We leave that for future work.

\paragraph{Asynchronous RL}
Although \sys mainly targets well-established synchronous RL algorithms, it can be easily extended to support asynchronous RL algorithms, e.g., one-step off-policy~\cite{zhong2025streamrl} or fully asynchronous~\cite{fu2025areal}, by not enforcing instances to use the latest weights in the rollout manager.

%% file: related.tex
\section{Related Work}

\paragraph{LLM training and serving on preemptible instances.}
Recent systems use preemptible instances to cut the cost of LLM pre-training and inference. For training, prior work achieves resilience via redundancy, migrating across parallelization strategies after preemptions and assuming spare replicas preserve lost model states~\cite{thorpe2023bamboo,duan2024parcae,jang2023oobleck,wu2024lazarus}. Under frequent preemptions, they often revert to checkpoint restarts and can stall when all replicas are lost, and they do not support fully non-redundant strategies such as FSDP~\cite{zhao2023fsdp}. Hence, preemptible-instance training remains inefficient and unstable. 

There is also a series of systems that leverage preemptible instances for online LLM serving~\cite{mao2025skyserve,miao2024spotserve}, but they focus primarily on optimizing service availability and latency SLOs. In contrast, RL workloads require optimizing the total execution time of each training step, which encompasses both interdependent rollout and training stages.

\paragraph{RL frameworks for LLMs.} A number of RL frameworks have been specifically designed for LLMs. NeMo-Aligner~\cite{shen2024nemo} and OpenRLHF~\cite{hu2024openrlhf} are among the earliest. They apply the disaggregated architecture and suffer from resource under-utilization due to serial dependencies between stages. Co-located frameworks like veRL~\cite{sheng2025hybridflow}, ReaL~\cite{mei2024realhf}, and RLHFuse~\cite{zhong2024optimizing} are hence proposed to improve resource efficiency. veRL~\cite{sheng2025hybridflow} combines single-controller and multi-controller paradigms to efficiently drive the execution of different stages. RLHFuse~\cite{zhong2024optimizing} introduces stage fusion to further improve GPU compute efficiency. Recently, the resource coupling issue of co-located frameworks has led to resurgent interests in the disaggregated architecture~\cite{zhong2025streamrl,han2025asyncflow,fu2025areal,he2025history}.
StreamRL~\cite{zhong2025streamrl} proposes a one-step off-policy training pipeline, where rollout uses stale weights that are one step behind. AReaL~\cite{fu2025areal} further relaxes this to fully asynchronous training. RhymeRL~\cite{he2025history} adopts the one-step off-policy paradigm and leverages speculative decoding to accelerate rollout. Notably, StreamRL~\cite{zhong2025streamrl} also proposes dynamically adjusting GPU resources allocated to training and rollout to elastically maintain balanced execution. However, it still assumes a fixed resource pool and cannot support preemptible resources, where resource availability is unpredictable.

%% file: conclusion.tex
\section{Conclusion}
In this paper, we present \sys, a hybrid RL framework that harvests preemptible GPU resources for high-throughput and cost-efficient RL on LLMs. \sys maintains a reserved (on-demand) training cluster while opportunistically offloading rollout workloads to preemptible instances. Through adaptive rollout offload with partial response seeding, \sys dynamically balances workloads between the training cluster and remote instances based on real-time resource availability. The pull-based weight transfer mechanism enables newly allocated instances to quickly join ongoing rollout, while token-level response collection minimizes preemption overhead and enables continuous load balancing. Experiments show \sys accelerates RL training by up to 1.97× while improving cost efficiency by up to 49\% compared to using only on-demand GPU resources, all while maintaining synchronous RL algorithms. 

\newpage

%% file: appendix.tex
\clearpage

\appendix
\section{Cloud Instance Cost} 
\label{sec:cloud_instance_cost}

\begin{figure*}[t]
\centering
\includegraphics[width=0.99\linewidth]{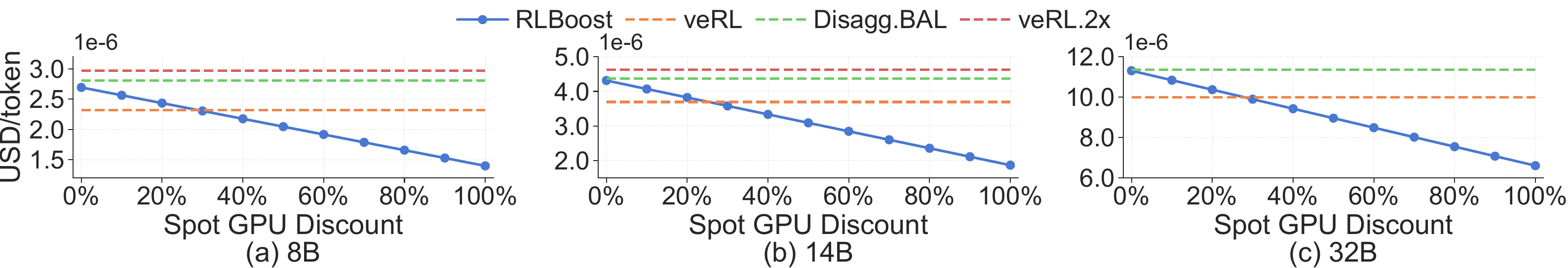}
\caption{{\textbf{[Cost sensitivity]:} The impact of spot GPU cost saving to the overall cost efficiency in \autoref{fig:overall-eval-bar}(d).}}
\label{fig:cost_sensitivity}
\end{figure*}

We calculate the average cost of instances with H100 across different regions on both AWS and GCP following~\cite{awspricing,gcloudpricing}. The results are listed in \autoref{tab:cloudpricing}. Both providers offer standard and spot provision options with large price gaps, and users can run \sys to boost RL throughput. 

\begin{table}[h]
\caption{Configurations of machines on public clouds for cost efficiency analysis.}
\label{tab:cloudpricing}
    \centering
    \small
    \resizebox{0.99\columnwidth}{!}
    {
    \begin{tabular}{ccccc}
        \toprule
        Provider & Machine Type & \#H100 & Provision & Cost/hour\\ 
        \midrule
        \multirow{3}{*}{GCP}  & \texttt{a3-highgpu-8g} & 8 & Standard & \$101.62\\
        \cmidrule{2-5}
         & \texttt{a3-highgpu-8g} & 8 & Spot & \$23.78\\ 
        \cmidrule{2-5} 
        & \texttt{a3-highgpu-2g}  & 2 & Spot & \$6.75 \\
        \midrule

        \multirow{3}{*}{AWS}  & \texttt{p5.48xlarge}  & 8 & Standard & \$65.96\\
        \cmidrule{2-5}
         & \texttt{p5.48xlarge}  & 8 & Spot & \$20.24\\ 
         \cmidrule{2-5}
         & \texttt{p5.4xlarge}  & 1 & Spot & \$1.76\\
        \bottomrule
    \end{tabular}
    }
\end{table}

A full \texttt{a3-highgpu-8g} is equipped with a 200 Gbps frontend NIC and four 200 Gbps backend NICs, while a \texttt{a3-highgpu-2g} can only access a 50 Gbps frontend vNIC~\cite{gcp_network}. On the other hand, a \texttt{p5.48xlarge} supports 3200 Gbps EFA network, while a \texttt{p5.4xlarge} only supports 100 Gbps EFA network~\cite{aws_network}. The limited bandwidth on fragmented instances makes them less feasible for distributed training, but
their high availability and affordable price make them a perfect fit for rollout.

For cost efficiency calculation, the average hourly cost of standard instance with 8 H100s on public clouds is 
$$\$(101.62 + 65.96)/2 \approx \$83.79,$$ 
and the average hourly cost of 2 spot H100s is 
$$\$(23.78/4 + 6.75+ 20.24/4 + 1.76\times 2)/4 \approx \$5.32$$

Figure~\ref{fig:cost_sensitivity} quantifies how spot GPU discounts affect \sys's cost efficiency across the three trace segments. We define spot GPU discount as the normalized per-GPU discount relative to on-demand: 0\% means equal price, and 100\% means effectively free GPUs (e.g., on-premise infrastructure). With no discount (0\%), \sys's cost efficiency lies between veRL and veRL.2x because \sys provisions additional (rollout) resources beyond veRL. With as little as \(\sim\)30\% savings, \sys improves cost efficiency while also increasing training throughput.

\section{Model Configuration} 
\label{sec:model_config}
In our evaluation, we use 8B/14B/32B models from Qwen3 family~\cite{qwen3technicalreport}. The models' details are listed in~\autoref{tab:eval_models_setup}.
\begin{table}[h]
\caption{Model configurations.}
\centering
\resizebox{0.99\columnwidth}{!}{
    \begin{tabular}{lcccc}
     \toprule
     & Layers & Q heads & K/V heads & Hidden size \\
    \midrule
    Qwen3-8B & 32 & 32 & 8 & 4096 \\
    \midrule
    Qwen3-14B & 48 & 48 & 8 & 5120 \\
    \midrule
    Qwen3-32B & 64 & 40 & 8 & 5120 \\
    \bottomrule
    \end{tabular}\label{tab:eval_models_setup}
}
\end{table}

\section{Characteristics of Preemptible Instance Traces} 
\label{sec:spot_trace}
Our preemptible instance trace is based on~\cite{thorpe2023bamboo}. To match our resource constraints, we randomly sampled 50\% of the instances from the original trace while preserving their individual allocation and preemption event histories. We evaluate \sys on three representative segments shown in \autoref{fig:overall-trace}. Their characteristics are listed in~\autoref{tab:spot_trace}. For best cost efficiency, we control the maximum number of running preemptible instances, $N_{\text{prem}}$, according to~\autoref{algo:adaptive_seeding}, instead of allocating all available instances. When an instance is preempted, i.e., \texttt{remove} event in the trace, the trace replayer will shut down the instance and immediately start a new one if available. 

\begin{table}[h]
\caption{Overview of the 3 segments of the spot instance trace.}
\centering
\resizebox{0.99\columnwidth}{!}{
    \begin{tabular}{lccc}
    \toprule
     Traces & Segment A & Segment B & Segment C \\
    \midrule
    Availability & High & Low & High \\
    \midrule
    Preemption Intensity & High & High & Low  \\
    \midrule
    \#Avg. Instances & 6.53 & 4.58 & 6.06 \\
    \midrule
    \#Allocations & 13 & 8 & 6  \\
    \midrule
    \#Preemptions & 8 & 9 & 2 \\
    \bottomrule
    \end{tabular}\label{tab:spot_trace}
}
\end{table}

\section{Impact of Weight Transfer Paradigm on Availability Spikes} 
\label{sec:weight_transfer_restart}

\begin{figure}[t]
\centering
\includegraphics[width=0.99\linewidth]{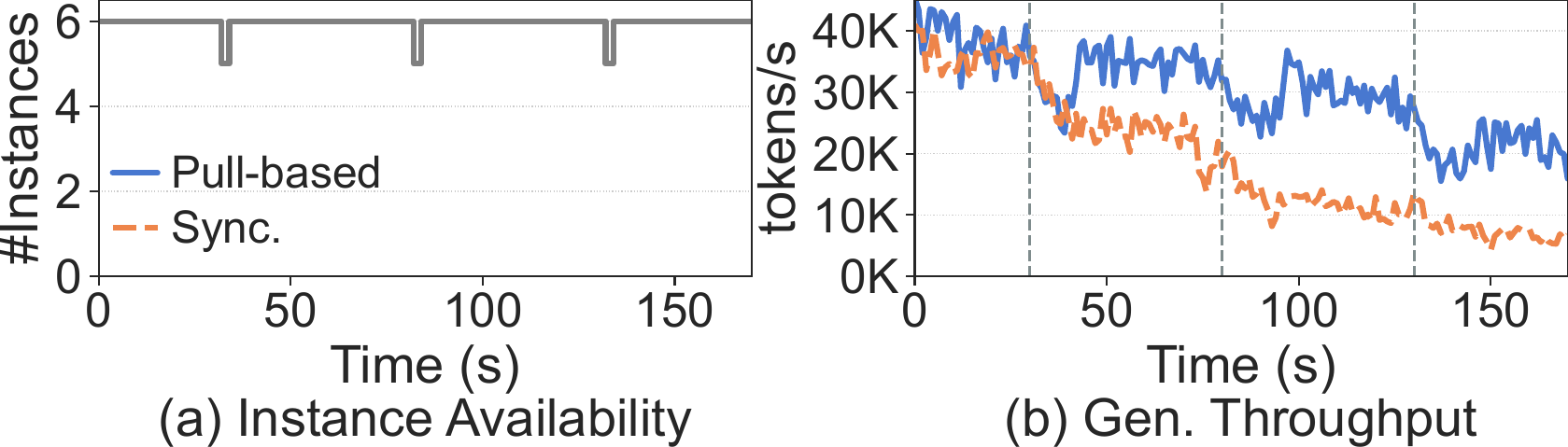}
\caption{\textbf{[Ablation study]:} Comparing pull-based and synchronized weight transfer as instances restart within a step. We use Qwen3-14B.}
\label{fig:ablation_weigh_transfer_restart}
\end{figure}

Besides allowing a newly available instance to quickly participate in the current step's rollout, the pull-based weight transfer agents also stabilize the throughput on availability spikes. As we observed in~\autoref{fig:overall-trace}, instances can be occasionally preempted, but a new one can be immediately allocated.  In~\autoref{fig:ablation_weigh_transfer_restart}, we construct a scenario where three rollout instances are preempted and restart consecutively within a step. Because the synchronous weight transfer logic updates weights between each step, a restarted instance cannot join the current step rollout, and the throughput drops accordingly. In contrast, with our pull-based weight transfer agents, the restarted instances immediately pull latest model weights from the agents and begin rollout, and we can observe the throughput quickly recovers.